\documentclass[%
longbibliography, 
 reprint,
superscriptaddress,
 amsmath,amssymb,
 aps,
]{revtex4-2}

\usepackage{physics}
\usepackage{graphicx}
\usepackage{amssymb}
\usepackage{amsmath}
\usepackage{epstopdf}
\usepackage{graphics}
\usepackage[caption=false,subrefformat=parens,labelformat=parens]{subfig}
\usepackage{float}
\usepackage{color}
\usepackage[colorlinks=true]{hyperref}
\usepackage{wasysym}
\usepackage{xcolor}
\usepackage{bm}

\begin{document}

\preprint{APS/123-QED}


\title{Minimizing readout-induced noise for early fault-tolerant quantum computers}

\author{Yunzhe Zheng}

 \email{dran.z@foxmail.com}
\affiliation{QunaSys Inc., Aqua Hakusan Building 9F, 1-13-7 Hakusan, Bunkyo, Tokyo 113-0001, Japan}%
\affiliation{QuTech, Delft University of Technology, Delft 2628 CJ, Netherlands}
\author{Keita Kanno}
\affiliation{QunaSys Inc., Aqua Hakusan Building 9F, 1-13-7 Hakusan, Bunkyo, Tokyo 113-0001, Japan}%

\date{\today}

\begin{abstract}

Quantum error correcting code can diagnose potential errors and correct them based on measured outcomes by leveraging syndrome measurement.  However, mid-circuit measurement has been technically challenging for early fault-tolerant quantum computers and the readout-induced noise acts as a main contributor to the logical infidelity. We present a different method for syndrome extraction, namely \textit{Generalized Syndrome Measurement},  that requires only a single-shot measurement on a single ancilla, while the canonical syndrome measurement requires multiple measurements to extract the eigenvalue for each stabilizer generator. As such, we can detect the error in the logical state with minimized readout-induced noise. By adopting our method as a pre-check routine for quantum error correcting cycles, we can significantly reduce the readout overhead, the idling time, and the logical error rate during syndrome measurement. We numerically analyze the performance of our protocol using Iceberg code and Steane code under realistic noise parameters based on superconducting hardware and demonstrate the advantage of our protocol in the near-term scenario.  As mid-circuit measurements are still error-prone for near-term quantum hardware, our method could boost the applications of early fault-tolerant quantum computing.
\end{abstract}

\maketitle


\section{Introduction}

Recent experimental progress has demonstrated powerful and scalable quantum computers with different hardware like superconducting qubits \cite{arute_quantum_2019, harrigan_quantum_2021, kjaergaard_demonstration_2022, google_quantum_ai_suppressing_2023, zhao_realization_2022, krinner_realizing_2022, marques_logical-qubit_2022}, neutral atoms \cite{bluvstein_quantum_2022,singh_dual-element_2022, graham_multi-qubit_2022, omran_generation_2019, levine_parallel_2019, evered_high-fidelity_2023, bluvstein_logical_2024} and trapped ions \cite{ringbauer_universal_2022, dumitrescu_dynamical_2022,ryan-anderson_realization_2021}. To release the full quantum power, practical quantum hardware needs to overcome various kinds of noise and neutralize their effect on the quantum states storing information. Quantum error correcting (QEC) codes provide a general framework to tackle noise by encoding a logical qubit into multiple physical qubits \cite{shor_scheme_1995,preskill_reliable_1998}. The extra 
Hilbert space allows the detection and correction of quantum errors, providing the possibility for large-scale quantum computing.

\begin{figure*}
    \centering
    \includegraphics[width = .9\linewidth]{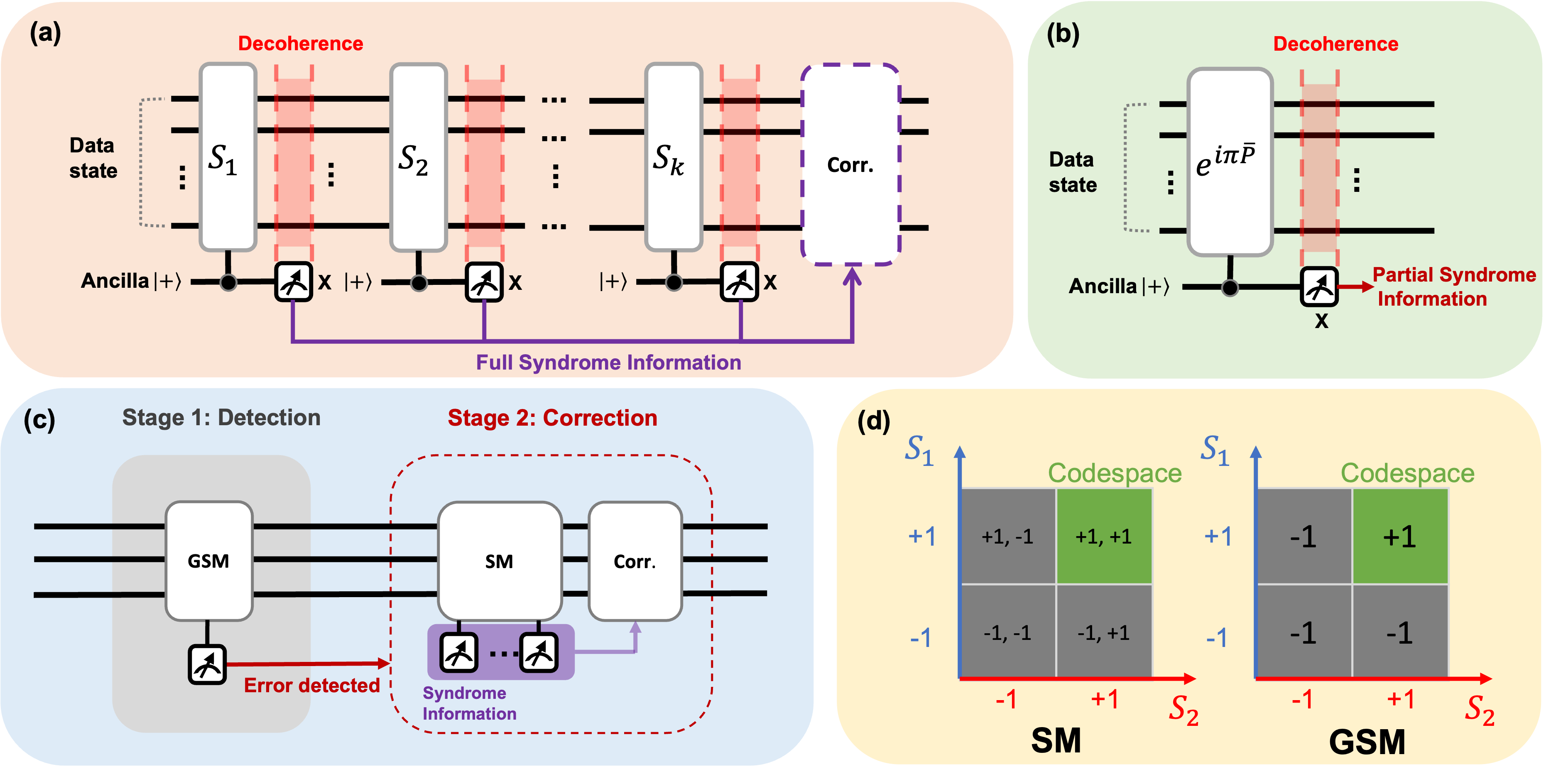}
    \caption{(a) Circuit scheme for canonical syndrome measurement (SM). A total number of $k$ (the number of stabilizer generators in the code) readouts are required to diagnose the error syndrome and apply proper correcting operations. While the ancilla qubit is being measured, all data qubits are subjected to decoherence due to idling. (b) Circuit scheme for (one-shot) generalized syndrome measurement (GSM). Only a single readout is needed to extract the partial information and detect the error, and the readout-induced idling error is minimized. (c) QEC cycle with generalized syndrome measurement. We divide the cycle into two stages: detection and correction. We first use the GSM method to check whether an error has occurred or not. If no error is detected, we skip the correction stage while ensuring that the logical state is in the codespace. If an error is detected, we follow the GSM routine by the canonical SM, extracting the full syndrome information and correcting the detected error. (d)  A two-stabilizer code example illustrates the difference between the SM method and the GSM method. The value within the box denotes the measured eigenvalue for $S_i$ (SM) or the code-space projector $\bar{P}$ (GSM).}
    \label{fig:1}
\end{figure*}

To diagnose physical errors in QEC codes, syndrome measurement (SM) is required to gain the necessary information for identifying both the location and type of errors. Let us consider a stabilizer code $Q$ which has a $k$-element stabilizer generator set $\{S_i\}$.  For each $S_i$, canonical syndrome measurement requires preparation of an ancilla state in $\ket{+}$, followed by applying a Pauli $S_i$ gate controlled by the ancilla and final measurement of the ancilla in the X basis (Fig.~\ref{fig:1}a). However, as mid-circuit readouts have been extremely error-prone for near-term quantum hardware, readout-induced noise significantly affects the fidelity of the logical state after the syndrome measurement. For superconducting qubits, the state-of-the-art readout duration ranges from hundreds of nanoseconds to several microseconds while the coherence time for scalable near-term hardware is tens of microseconds \cite{google_quantum_ai_suppressing_2023, zhao_realization_2022}. When the ancilla qubit is being measured out, all data qubits are idling and therefore suffer from decoherence noise. Besides, the state-of-the-art readout assignment error is on the scale of $0.01$, which cannot be mitigated if no repeated readouts are performed. For neutral atom arrays, onsite mid-circuit readouts are practically challenging and the ancilla qubit to be measured must be moved to a separate zone by coherent transportation to avoid unintended interaction on other physical qubits \cite{bluvstein_quantum_2022, bluvstein_logical_2024, evered_high-fidelity_2023}, which adds time overhead and might be susceptible to unexpected errors. 

Exploiting redundant ancilla qubits and measuring different stabilizers in parallel may alleviate the readout-induced decoherence, but it poses a stricter requirement for qubit control and connectivity and may bring further space overhead for near-term quantum hardware. The direct readout error, e.g. assignment error and crosstalk, is still present even if every stabilizer is measured simultaneously with separate ancilla qubits. Therefore, we consider it favorable to reduce the number of readouts for better performance on early fault-tolerant quantum computers. Notably, there has been work regarding measurement-free QEC protocols that could achieve fault-tolerant quantum computing without ancilla readout \cite{nebendahl_optimal_2009, paz-silva_fault_2010,crow_improved_2016,heusen_measurement-free_2024}. However, these protocols require the ability to reset the qubits, which is generally as noisy as readout in the near-term quantum hardware \cite{evered_high-fidelity_2023, zhou_rapid_2021, sunada_fast_2022}. Therefore, it in principle just transfers the overhead from readout to resetting and only works for limited quantum hardware where resetting is much less error-prone than readout.

In this work, we propose a different method to extract syndrome information, namely \textit{Generalized Syndrome Measurement} (GSM), that requires only single measurement on a single ancilla regardless of the number of stabilizer generators. As such, our protocol would minimize the time of readout for syndrome check. Specifically, we check the encoded states by directly measuring the eigenvalue of the code-space projector instead of the eigenvalue of each stabilizer generator, which allows us to check whether an error occurred or not at a single-shot readout overhead. Further, we propose to exploit the GSM method as a pre-check routine for canonical syndrome measurement in QEC cycles, which may render the following canonical routine unnecessary depending on the GSM readout outcome. As a result, our method can reduce the average number of readouts for quantum error correction, given that the input logical state only suffers from low level of noise. We numerically demonstrate the performance of our protocol with Iceberg code \cite{self_protecting_2024} and Steane code \cite{steane_multiple-particle_1997}, and show that our method outperforms the canonical method in the practical noisy scenario. As long as mid-circuit readouts are still a main infidelity source for quantum hardware, we believe our method can reduce the impact of readout-induced noise and enhance the performance of quantum applications in early fault-tolerant quantum computing.

\section{Protocol Description}

We first introduce the idea of the GSM method. For a QEC code $Q$ with stabilizer generator $\{S_i\}$($i=1,2,...,k$), each stabilizer generator $S_i$ has an associate projector $P_i = (I+S_i)/2$. We can define the code-space projector of $Q$ 
\begin{equation}\label{eq:P}
    \bar{P} = \prod^{k}_{i=1} P_i = \frac{1}{2^k}\prod^k_{i=1} (I+S_i).
\end{equation} 
Instead of measuring each stabilizer individually, we aim to detect whether a given state is in the codespace or not by measuring the eigenvalue of $\bar{P}$ using controlled unitary gates $e^{ i\pi\bar{P}}$. As shown in Fig.~\ref{fig:1}b, We start with the noisy input logical state $\ket{\psi}$ and an ancilla initiated at $\ket{+}_a$ 
\begin{equation}
    \frac{1}{\sqrt{2}}(\ket{0}_a\ket{\psi}+\ket{1}_a \ket{\psi}).
\end{equation}
Next, we apply a control-$e^{i\pi \bar{P}}$ gate triggered by the ancilla state $\ket{1}_a$, and the pre-measurement state becomes 
\begin{equation}
    \frac{1}{\sqrt{2}}(\ket{0}_a\ket{\psi}+\ket{1}_a e^{i\pi \bar{P}}\ket{\psi}).
\end{equation}
Rewriting the ancilla state in the X basis, we have
\begin{equation}
    (\ket{+}_a (I-\bar{P})\ket{\psi}+\ket{-}_a \bar{P} \ket{\psi}),
\end{equation}
where we used the fact that $\bar{P}^2 = \bar{P}$ and $e^{i\beta\bar{P}} = I + (e^{i\beta} - 1)\bar{P}$ for any angle $\beta$. If we measure the ancilla qubit to be $\ket{-}_a$, the logical state will be projected to the state proportional to $\bar{P}\ket{\psi}$, which lies in the code space of $Q$, and we assure there is no detectable error within the ability of code $Q$. Otherwise, the post-measurement state lies outside of the codespace, implying that at least one detectable error has occurred in the logical state. 

Compared with the GSM method, the canonical SM method projects the input state to the codespace (+1 eigenspace of all $S_i$) by measuring the ancilla qubits sequentially. As each stabilizer $S_i$ requires an ancilla readout, at least $k$ readouts are required to tell whether the state is in the codespace or not and the readout-induced idling time will be $kt_{m}$ given $t_{m}$ is the time for a physical readout. The direct readout errors, like assignment errors and crosstalk errors, may also affect the fidelity of the logical state by $k$ times. In the GSM method, we require only a single measurement to extract partial syndrome information and diagnose if an error exists, at the cost of losing the specific information of which kind of error exists. Therefore, the GSM method can take the place of the SM method for quantum error detection (QED), where we post-select logical states based on readout outcomes. Fig.~\ref{fig:1}d provides an illustrative example of a code containing two stabilizer generators. For the purpose of QEC, the GSM method cannot completely replace the canonical SM method as all syndrome information is necessary. However, the GSM method can still bring benefits for QEC cycles: In a fault-tolerant quantum computing regime, The logical error rate of the input state is generally very low to ensure that we are under the noise threshold where logical qubits would outperform physical qubits. Therefore, the main purpose of a QEC cycle is to assure that no error has occurred in the input logical state, rather than revealing the specific information about the presented errors. Based on this assumption, we can integrate the GSM method in QEC cycles in an adaptive way and divide a cycle into two stages as presented in Fig.~\ref{fig:1}c: In the first stage, we detect if there is an error in the logical state using the GSM method. If no error is detected (in a more probable case), we assure the logical state is in the codespace, skip the next stage, and finish this QEC cycle. If at least an error is indeed detected (in a less probable case), we follow up with the canonical SM method to extract the full syndrome information and correct all detected errors. As long as the input state has sufficiently high fidelity, we can lower the average time of ancilla readouts by introducing the GSM method in such a combined way. For example, if the input state (encoded in a $k$-stabilizer code) is the ideal state with $99\%$ probability or a fully erroneous state with $1\%$ probability, the average readout-induced idling time for a noiseless QEC cycle with the GSM method can be approximated by $[0.99\cdot1+0.01\cdot(1+k)]t_m = (1+0.01k)t_m$, which is much shorter than the time $kt_m$ for the canonical method even if $k=2$. 

Notably, the idea of adaptive syndrome measurement has already been used for Shor-style error correction \cite{delfosse_short_2020, tansuwannont_adaptive_2023}, but we are using it here in a different way: The first measurement is for the whole codespace projector, and the following measurements are for each stabilizer generator. Our method could also be further made in Shor-style fault-tolerant way (see Appendix \ref{app-faulttolerance}).   As mid-circuit readouts are still very error-prone for current large-scale quantum hardware, we expect our method could significantly reduce the noise originating from readouts and benefit early fault-tolerant quantum applications. 

In addition to the one-shot GSM method shown in Fig.~\ref{fig:1}(b) where we just use one ancilla readout, one can also use a few more ancilla readouts (but still less than the canonical SM method) for a more reduced gate complexity and relaxed connectivity requirement. This trade-off can be achieved by splitting the code-space projector $\bar{P}$ into several subprojectors $\{\bar{P}_1, \bar{P}_2,..., \bar{P}_m \}(m<k)$ such that
\begin{equation}
        \bar{P} = \prod^{m}_{i=1} \bar{P}_i  
\end{equation}
and 
\begin{equation}
    \bar{P}_i = \prod_{\{j\}_i} P_j,
\end{equation}
where $\{j\}_i$ is an index set associated with each subprojector and $\bigcup^{m}_{i=1} \{j\}_i = \{x\in \mathbb{Z} | 0\leq x\leq k \}$. By measuring these subprojectors $\bar{P}_i$ separately, we can achieve an $m$-shot GSM protocol where the total readout time is still less than the canonical SM method.  For example, a four-stabilizer code (e.g. the five-qubit $[[5, 1, 3]]$ code) with ${S_1, S_2, S_3, S_4}$ can exploit a two-shot GSM by splitting the general projector into two subprojectors, one containing $S_1$ and $S_2$ while the other containing $S_3$ and $S_4$. We give a more detailed explanation in the Appendix \ref{app_m_shot}.

\section{Gate implementation of $e^{i\pi\bar{P}}$}

\begin{figure}
    \centering
    \includegraphics[width = \linewidth]{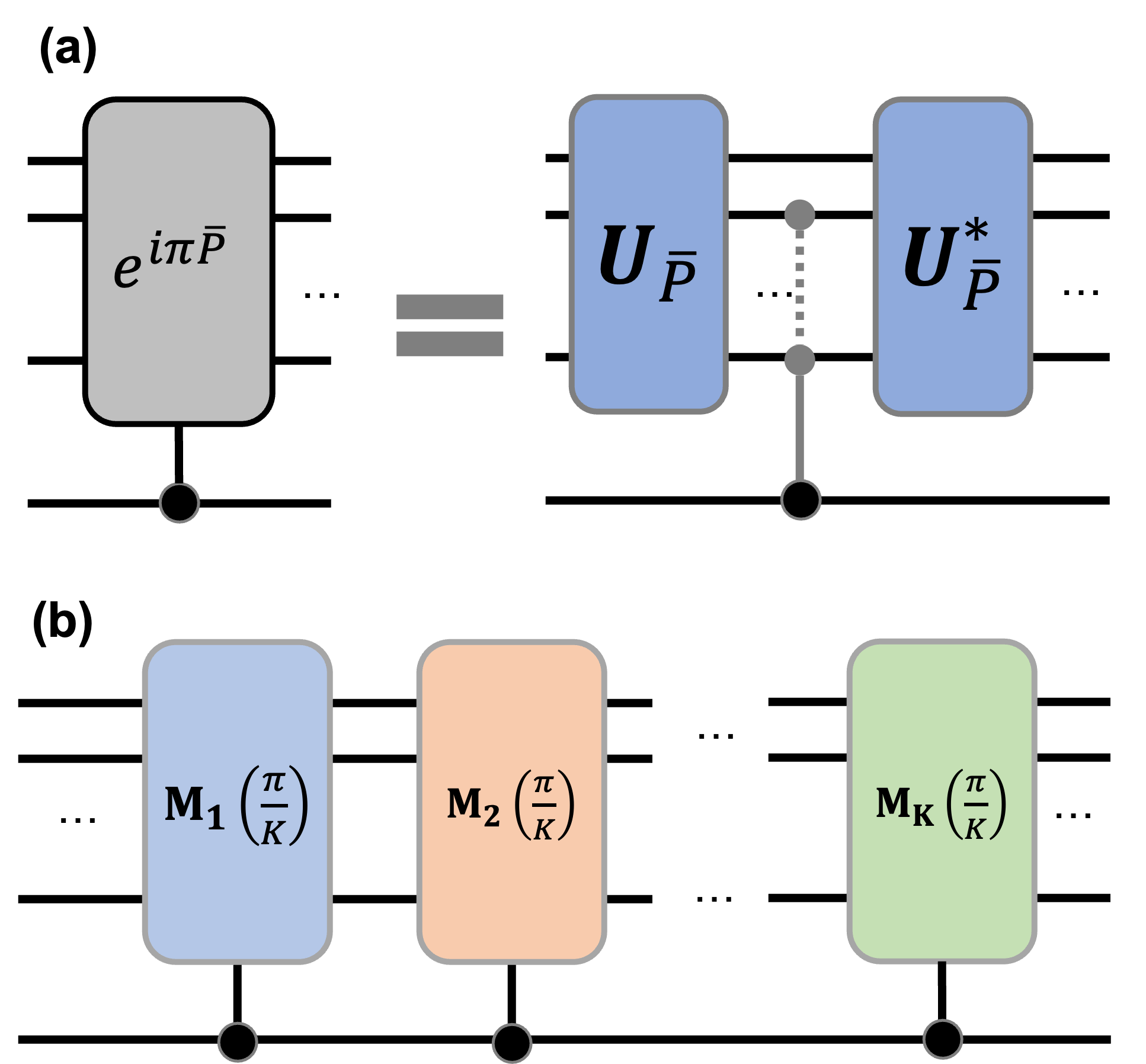}
    \caption{Gate Implementation of $e^{i\pi\bar{P}}$. (a) Sandwich implementation. $U_{\bar{P}}$ is the circuit decoding the code with projector $\bar{P}$ back to the trivial code state, and there is a multi-qubit control phase gate sandwiched in the middle between $U_{\bar{P}}$ and its conjugate $U^*_{\bar{P}}$. (b) The $e^{i\pi\bar{P}}$ can also be implemented by a sequence of control multi-qubit Pauli rotation gates. }
    \label{fig:gate}
\end{figure}

\begin{figure*}[t]
    \centering
    \includegraphics[width=\linewidth]{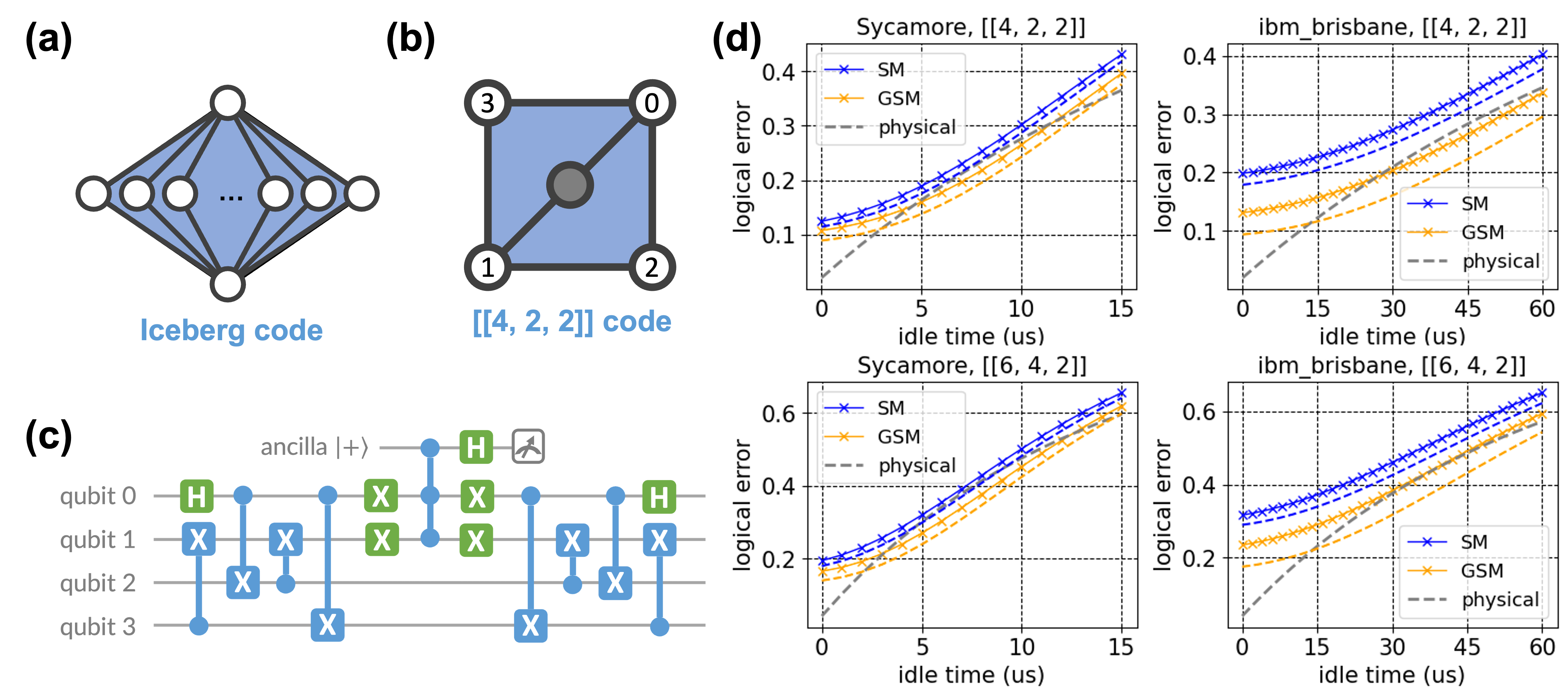}
    \caption{(a) Planar connectivity structure of Iceberg codes. All physical qubits are placed in an iceberg-like shape and the bottom/up qubit is connected to every other qubit. (b) The $[[4, 2, 2]]$ code with the required ancilla connectivity for GSM. (c) Decomposed circuit for GSM on the $[[4, 2, 2]]$ code. (d) Numerical simulation of four-qubit and six-qubit Iceberg code, with noise parameters of \textit{Sycamore} \cite{google_quantum_ai_suppressing_2023} and \textit{ibm\_brisbane} processor \cite{noauthor_ibm_nodate}. The dashed curve indicates error rate when the gate control is perfect and all gate noise is gate-induced idling noise.}
    \label{fig:iceberg}
\end{figure*}
In the GSM method, we demand the implementation for the controlled projector gates $e^{ i\pi\bar{P}}$, which is a highly non-local unitary and requires to be decomposed into elementary gates to be implemented in near-term quantum hardware. Here, we provide two different methods of implementation.

In the first method of decomposition, we expand the expression of $\bar{P}$ in Eq.~(\ref{eq:P}) into a sum of stabilizer operators $M_i$:
\begin{equation}
    \bar{P} = \frac{1}{2^k}\prod^k_{i=1} (I+S_i)=\frac{1}{K}\sum^{K}_{i=1}M_i,
\end{equation}
where $K=2^k$ for simplicity. As $S_i$ commutes with each other, $M_i$ is also commutable with each other. Thus, an $e^{i\pi\bar{P}}$ gate can be written as
\begin{equation}
    e^{i\pi \bar{P}} = \prod^K_{i=1}e^{i\pi M_i/K} = \prod^K_{i=1} \mathbf{M_i}(\pi/K),
\end{equation}
where $\mathbf{M_i}(\theta)=e^{i\theta M_i} $ are multi-qubit Pauli phase gates with angle $\theta$. The controlled projector gates are therefore rearranged in the gate sequences as shown in Fig.~\ref{fig:gate}b. As the identity gate $I$ is also among $\mathbf{M_i}$, there are $K-1$ non-trivial gates that need to be implemented in practice. This kind of implementation is particularly suitable for systems with globally tunable Hamiltonian, like trapped ions 
 \cite{wetering_constructing_2021} or neutral atoms \cite{jandura_time-optimal_2022}, but can also be decomposed into local Clifford and single-qubit phase gates \cite{clinton_towards_2024}. Notably, although the number of multi-qubit Pauli gates grows exponentially with $k$, the total physical Hamiltonian evolution time is bounded as the rotation angle for each $\mathbf{M}_i$ also decays exponentially with $k$.

In the second method, we can construct the controlled $e^{i\pi\bar{P}}$ using a multi-qubit control phase gate sandwiched by a Clifford unitary $U_{\bar{P}}$ and its conjugate as shown in Fig.~\ref{fig:gate}a. The unitary $U_{\bar{P}}$ always exists as any $[[n, k, d]]$ stabilizer codes can be encoded by a Clifford unitary from the trivial code state
\begin{equation}
    \ket{\psi}_{data}\otimes \ket{1}_{k+1}\otimes... \otimes \ket{1}_{n},
\end{equation}
where $Z_{k+1},...,Z_{n}$ is the stabilizer generator of the trivial code \cite{cleve_efficient_1997}. Therefore, only logical states in the codespace will be decoded into the trivial code state with all non-data qubits remaining in the $\ket{1}$. A multi-qubit control phase gate will therefore gain the necessary information to detect if the logical state has left the codespace. For arbitrary stabilizer codes, the general decomposition of $U_{\bar{P}}$ can be found by Gottesman's algorithm \cite{gottesman_stabilizer_1997}.

\section{Error detection with Iceberg code}

$[[2m+2, 2m, 2]]$ Iceberg code is named by its planar connectivity requirement (Fig.~\ref{fig:iceberg}a) and has stabilizer generators $\{X^{\otimes 2m+2}, Z^{\otimes 2m+2} \}$. Although this kind of distance-2 codes cannot correct any errors but only detect a single error, the large encoding rate and loose connectivity requirements render great practical interest in near-term algorithm application \cite{self_protecting_2024, yamamoto_demonstrating_2024, urbanek_error_2020}. To demonstrate the performance of the GSM method, we numerically simulate the state in density matrix formalism using \textit{qiskit.quantum\_info} \cite{qiskit_contributors_qiskit_2023}. We take the noise parameters from the latest \textit{Sycamore} Processor \cite{google_quantum_ai_suppressing_2023} (with tunable couplers) and the \textit{ibm\_brisbane} Processor \cite{noauthor_ibm_nodate} (with fixed-frequency couplers). We consider initialization noise, gate noise, idling decoherence, and readout assignment error. Particularly, we model the two-qubit gate noise as a combination of a depolarizing channel and an idling decoherence channel with duration equal to the gate time to fully reflect the effect of gate time on the performance. We discuss our simulation in more details in the Method section. Although we only simulate the case when $m=1$ and $m=2$, we expect larger Iceberg code should exhibit similar numerical behavior as the number of stabilizer generators doesn't increase with the code size.

 We first use noisy gates to prepare the physical qubits in the GHZ state 
\begin{equation}
    \frac{1}{2}(\ket{0}^{\otimes 2m+2}+\ket{1}^{\otimes 2m+2}),
\end{equation}
which is the logical $\ket{\overline{+}}^{\otimes{2}}$ for $[[4, 2, 2]]$ code and $\ket{\overline{+}}^{\otimes{4}}$ for $[[6, 4, 2]]$ code up to local transversal Hadamard. Next, we apply a varying period of idling noise on every physical qubit to mimic realistic noise. We then separately use the canonical SM or the GSM method to detect the error and post-select the state without triggered syndrome. We adopt the sandwich implementation of the control projector gate and the decomposed circuit for the four-qubit code is shown in Fig.~\ref{fig:iceberg}c, which requires the implementation of physical $CCZ$ gates or further decomposition. This implementation suits better for quantum hardware with less flexible qubit connectivity. The general implementation of the GSM method on the Iceberg code is discussed in Appendix \ref{appendix-gate}. We plot the logical error rate against the idling time in Fig.~\ref{fig:iceberg}d as well as the raw physical state prepared in $\ket{+}^{\otimes 2m}$ under idling noise for reference. We also plot the logical error rate when the gate noise is fully dominated by gate-induced decoherence as the dashed curve with the same colors.

For all four simulations shown in Fig.~\ref{fig:iceberg}d, the GSM method outperforms the canonical SM method by a significant gap. The amount of error rate improvement for the GSM method is barely affected by the idling time applied to the input logical state. In the case of state-of-the-art quantum hardware, the canonical SM method doesn't even have a better performance than the unencoded physical state, i.e. in a regime where the logical error rate is lower than the raw error rate. Nevertheless, the GSM method can achieve break-even and outperform the unencoded physical state after a certain idling time. The advantage of GSM is even larger when the control is perfect and gate noise is fully dominated by the gate-induced decoherence. As the gate control techniques are consistently improving while the gate time is limited by physical constraints in many hardware implementations, such as the state leakage in superconducting qubits  \cite{arute_quantum_2019} or the operating laser power \cite{evered_high-fidelity_2023} for neutral atoms, we anticipate the gate noise in the future quantum hardware to be more dominated by gate-induced decoherence and our protocol could therefore bring further advantage over the canonical protocol.

\section{Error correction with Steane code}

\begin{figure*}[t]
    \centering
    \includegraphics[width=\linewidth]{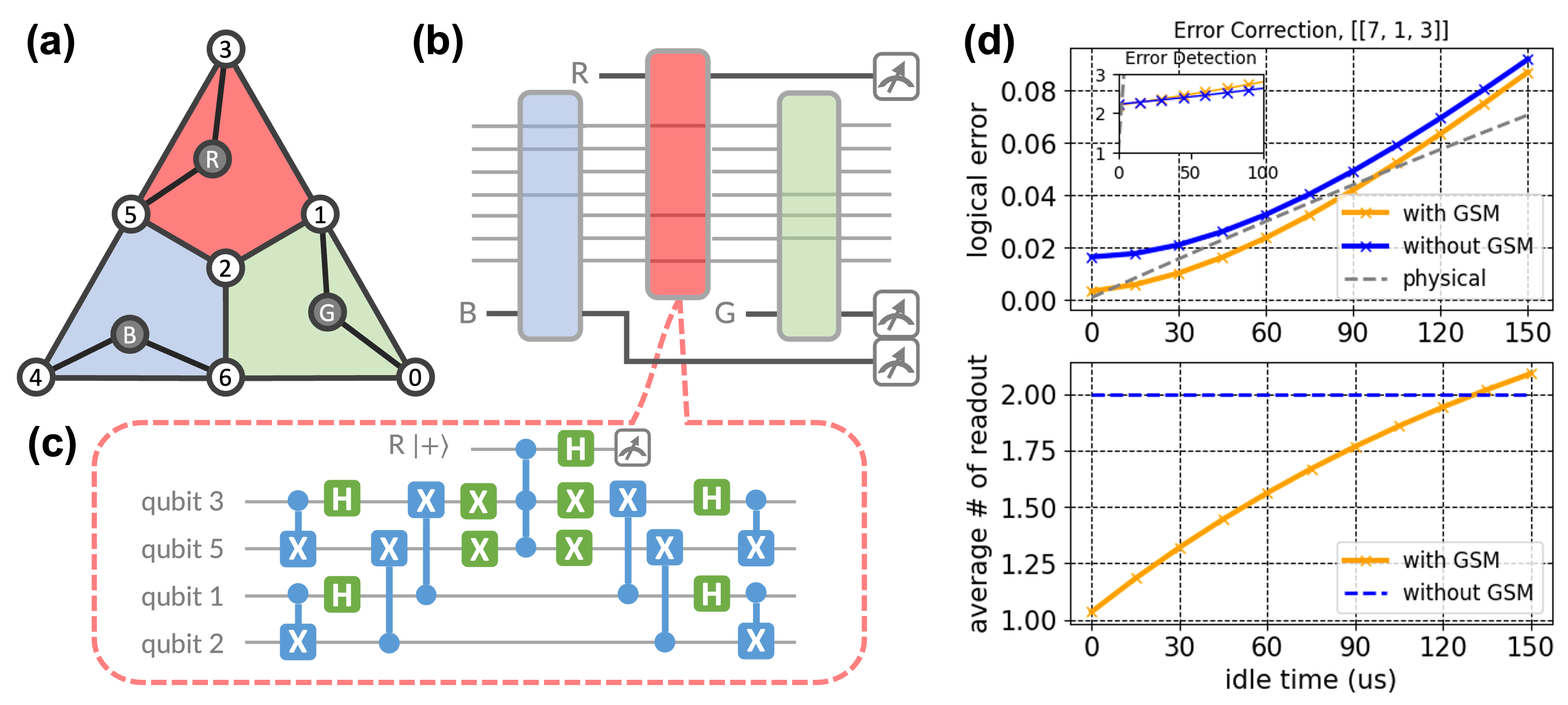}
    \caption{(a) Planar connectivity structure of Steane code for the GSM method. On each color plaquette, the ancilla qubit is only required to connect two neighboring physical qubits on the edge. (b) Circuit scheme for three-shot GSM on Steane code. The check on each plaquette is performed sequentially but the ancilla qubits are measured in the same time. (c) Detailed implementation for plaquette check. (d) Logical error rate under idling noise after performing QEC cycle with (without) GSM and the corresponding number of average readout required. The inset figure is the logical error rate after just performing post-selection with quantum error detection (y axis label in the unit of 0.001).  }
    \label{fig:steane}
\end{figure*}

Although quantum error detection can bring benefits for very near-term scenario, the post-selection routine introduces sampling overhead and quantum error correction is needed from a scalable viewpoint. To demonstrate the advantage of the GSM method for quantum error correction, we choose the $[[7, 1, 3]]$ Steane code to numerically compare the logical performance of the GSM with canonical SM method. The Steane code is the smallest two-dimensional color code that is able to correct a single error, and has weight-four stabilizer $XXXX$ and $ZZZZ$ on each of the three color plaquettes shown in Fig.~\ref{fig:steane}a. Rather than using the one-shot GSM method that projects the logical state onto the code space with a single readout, we choose to use the three-shot GSM to alleviate the connectivity requirement for planar layout of physical qubits. Specifically, we use three ancilla qubits, each to measure the projector of stabilizer $XXXX$ and $ZZZZ$ on its color plaquette. As all the three ancilla can be measured out simultaneously, the three-shot GSM method can still reduce the readout time as the canonical SM require two readouts on each ancilla qubit. In principle, we can use the circuit implementation for four-qubit Iceberg code shown in Fig.~\ref{fig:iceberg}c for each plaquette check as the stabilizer set is the same. However, we use an alternative implementation (Fig.~\ref{fig:steane}c) here to avoid the excessive connection on any single qubits such that each physical qubit is connected to at most four other physical qubits. In such a way, our GSM method should be exploited in state-of-the-art superconducting hardware without any further physical modification.

We use a projective noise parameter set for Steane code simulation because the state-of-the-art quantum hardware is still too noisy for logical qubits to outperform physical qubits see Appendix \ref{app-stateofart}. We initialize the logical qubit in $\ket{\bar{+}}$ with noisy circuits, apply idling noise to every physical qubit, and perform quantum error correction on the noisy logical state. We compare the performance of the QEC cycle with GSM (Fig.~\ref{fig:1}c) and without GSM (Fig.~\ref{fig:1}a), and plot the logical error rate and the average number of readout in Fig.~\ref{fig:steane}d. More simulation details can be found in the Method section as well.

With the help of the GSM method,  the logical error rate after a QEC cycle is significantly reduced when the applied idling noise is low. We attribute the advantage of GSM to the reduced average time of readout: When there is little idling noise, the input state is nearly perfect and the pre-check routine can assure the correctness of the input state with only a single-shot readout, leading the average number of readout close to one. In contrast, the canonical SM method always needs to measure twice and the readouts itself introduce additional noise to the logical state. When the idling noise increases and the input state becomes noisier, the average time of readout is increased as the GSM routine has more chance to detect an error and trigger the following correction process. However, even when the average time of readout is the same for QEC with or without GSM, we still see non-zero advantage in the QEC process with GSM. We attribute this advantage to the higher robustness of the GSM method on false-positive readout assignment errors: When the logical state is in the codespace but the readout outcome wrongly indicates there is an error in the detection stage, the QEC cycle with the GSM method will not be affected as the following correction routine will not detect any error and assure that the logical state is in the codespace.

\section{Discussion}

A natural question arise from our work is: Does there exist a range where the GSM method doesn't bring advantage on the logical fidelity? As the GSM method trades additional gate overhead for reduced readout overhead, the ratio between readout time and gate time can be used as the criterion. We numerically study the effect of readout time and gate time on the relative performance of the SM method and the GSM method in the supplemental material in Appendix \ref{app-parachoice} and find out the parameter of practical superconducting quantum hardware falls in the regime where the GSM method is advantageous. Therefore, unless there is a technical breakthrough that shortened the physical readout time drastically, we believe the GSM method will bring practical advantage for near-term quantum applications.

Besides using the GSM method as a pre-check routine, fault-tolerant quantum computing can also benefit from GSM by exploit concatenated codes and treat detected physical errors as logical erasure errors \cite{li_concatenation_2023}, which doesn't require explicit quantum error correction in the inner code but only need to correct logical erasure in the outer loss-tolerant codes (see Appendix \ref{app-logicalerasure}). In such a case, our GSM method can be exploited to improve the robustness against readout-induced error. 

Notably, as the $e^{i\pi\bar{P}}$ gate used in the GSM method needs to be decomposed into elementary gates, local errors might propagate to other physical qubits and produce correlated errors. However, as long as readout-induced idling error, which is a global noise channel, still serves as the main noise contributor in the near-term scenario, the correlated errors won't weigh over the readout-induced errors and pose a critical threat to the validness of our proposed method. Moreover, we may also use the idea of Shor-style syndrome measurement, which rely on pre-entangled ancilla qubits and native multi-qubit Pauli gates to achieve gate-level fault tolerance at the cost of more ancilla qubits and time overhead. (see Appendix \ref{app-faulttolerance})

Although we only simulate the performance of the GSM method using noise parameters from superconducting hardware, the GSM method can generally be applied to various kinds of quantum hardware where mid-circuit readouts are error-prone. For neutral atoms, state-of-the-art readout time is on the scale of tens of milliseconds and the best reachable coherence time is on the scale of seconds \cite{singh_mid-circuit_2023, bluvstein_quantum_2022}. In particular, mid-circuit measurements for neutral atoms rely on coherently transporting the atoms to a separate area as shown in \cite{bluvstein_quantum_2022}, or need to shelve the data qubits to magnetic-sensitive states with reduced coherence time \cite{graham_midcircuit_2023}. Our protocol can therefore be employed to provide advantages for scalable neutral atoms quantum computers as well.

\smallskip

\section*{Acknowledgement}

We thank Johannes Borregaard, Barbara Terhal, Christian Andersen, Masaya Kohda, Yuya O. Nakagawa and Allen Zang for insightful discussions. We thank Kah Jen (Bernard) Wo, Yanwu Gu, Senrui Chen, Xiaoyu Liu, and Dingshan Liu for proofreading the manuscript. 

\appendix

\section{Simulation details}

All of our numerical results are obtained from simulation using \textit{qiskit.quantum\_info} \cite{qiskit_contributors_qiskit_2023}. The quantum states are simulated in density matrix formalism and all error channels are implemented as Kraus superoperators. All physical qubits are prepared in $(1-\epsilon_{init})\ket{0}\bra{0}+ \epsilon_{init}\ket{1}\bra{1}$ where $\epsilon_{init}$ is the initialization error. For single-qubit (1Q) gates, we only consider the control error and model the noise by following a depolarizing channel after the ideal gate operation. For two-qubit (2Q) gates, we consider both the control error and the gate-induced decoherence.  We follow a two-qubit depolarizing channel after the ideal 2Q gate operation, and a global idling channel to all physical qubits after a layer of 2Q gates. In such a way, we ensure the total infidelity of a 2Q gate under control noise and the idling noise equals to the median error rate reported in \cite{google_quantum_ai_suppressing_2023} and \cite{noauthor_ibm_nodate}. For the Toffoli gate used in the GSM method, we model the noise by follow a three-qubit depolarizing channel with double error rate of 2Q gate and double gate-induced idling duration. Although most large-scale quantum hardware by far doesn't support native Toffoli gates and require further decomposition into five two-qubit gates, we are taking an optimistic consideration as native multi-qubit gates have been demonstrated on various kinds of quantum hardware \cite{evered_high-fidelity_2023, kim_high-fidelity_2022}.  For noisy readout, we model it by inserting an idling channel whose duration is equal to readout time before measurement. We also considered the readout (assignment) error that read $\ket{0}$ as $\ket{1}$ and vice versa. The specific noise parameters can be found in Table.~\ref{table_parameter}.

\begin{table}
\begin{tabular}{||c c c c||} 
 \hline
 - & \textit{Sycamore} & \textit{ibm\_brisbane} & projective \\ [0.5ex] 
 \hline\hline
 $T_1$($\mu s$) & 20 & 217 & 1000 \\ 
 \hline
 $T_2$($\mu s$) & 30 & 130 & 1000 \\
 \hline
 readout time($ns$) & 660* & 4000 & 200 \\
 \hline
 1Q gate error & 1e-3 & 2.2e-4 & 0 \\
 \hline
 2Q gate time$(ns)$ & 34 & 600 & 20 \\
 \hline
 2Q gate error & 3e-3 & 7.5e-3 & 1e-4 \\
 \hline
 initialize error & 1e-2 & 1e-2 & 1e-3 \\
 \hline
 readout error & 2e-2 & 1e-2 & 1e-3 \\ [1ex]
 \hline
\end{tabular}
*including time for qubit resetting
\caption{Noise parameter for numerical simulation}
\label{table_parameter}

\end{table}

For the Iceberg code simulation, we initialize the logical state by preparing a multi-qubit GHZ state. We start with a physical qubit in $\ket{+}$ and all others in $\ket{0}$ and apply CNOT gates sequentially. The depth of initialization for four(six)-qubit code is 2(3) and can be achieved with the iceberg-like connectivity. We assume only one available ancilla, which connects to every other physical qubit in the SM method but only require connection to two physical qubits in the GSM method. The gate depth of a cycle of the GSM method is 4(8) plus a Toffoli depth for the 4(6)-qubit Iceberg code. Comparatively, the depth for the SM method is 8(12).

For the Steane code simulation, we prepare the logical state by applying nine CZ gates (depth 3) on physical qubits all prepared in $\ket{+}$ and Hadamard gates on four of them \cite{bluvstein_quantum_2022}. We use three ancilla qubits, one for each color plaquette. In the QEC cycle without the GSM method, we first perform the X check on each plaquette, then the Z check on each plaquette. In the QEC cycle with the GSM method, we apply the partial projector gate on each ancilla sequentially (Fig.~\ref{fig:steane}b) before reading out all three ancilla qubits. The gate depth for the GSM method is 12 plus 3 Toffoli depth. In contrast, the canonical SM has 8 depth of 2Q gates.  

\section{$m$-shot Generalized Syndrome Measurement with subprojectors}
\label{app_m_shot}
\begin{figure}[b]
    \centering
    \includegraphics[width=\linewidth]{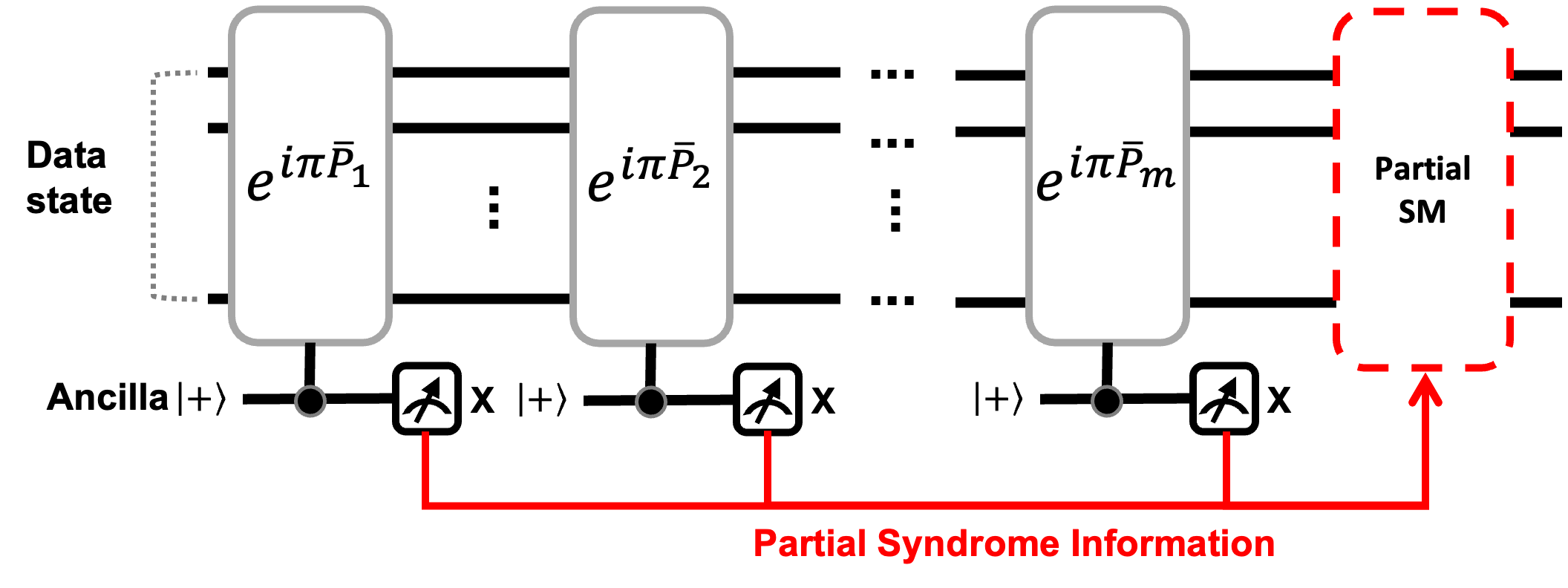}
    \caption{Circuit scheme for $m$-shot generalized syndrome measurement.}
    \label{fig:subprojector}
\end{figure}

In the main text, we introduced the generalized syndrome measurement (GSM) which only requires a single readout to detect the error. The GSM method requires the ability to implement controlled projector gates $e^{ i\pi \bar{P}}$, which could bring large computational overhead when decomposed into elementary gates and the connectivity requirement could be very strict. Here we introduce the modified method by exploiting a bit more ancilla qubits, which could help reach a smooth trade-off between the gate complexity/connectivity and the readout complexity in the GSM method. 

For code $Q$ with stabilizer generator $\{S_i\}$, the general projector for one-shot GSM method is 
\begin{equation}
    \bar{P} = \prod^k_i P_i = \frac{1}{2^k}\prod^k_i (I+S_i) = \frac{1}{K}\sum^{K}_i M_i.
\end{equation}
As we showed in the main text, we can use a control-$e^{i \pi \bar{P}}$ gate and a single-shot measurement on the ancilla qubit to act the projector on the noisy state nondeterministically. Instead of using only one projector, we can group several subprojectors $\bar{P}_i$ such that their product is still $\bar{P}$:
\begin{equation}
    \bar{P} = \prod^m_i \bar{P}_i,
\end{equation}
where
\begin{equation}
    \bar{P}_i = \prod_{\{j\}_i } P_{j}
\end{equation}
and $\{j\}_i$ is an index set associated with each subprojector and $\bigcup^{m}_{i=1} \{j\}_i = \{x\in \mathbb{Z} | 0\leq x\leq k \}$. As we have $m$ subprojectors, we need $m$ ($1\leq m \leq k$) readouts for partial syndrome extraction here. We can use the circuit show in Fig.~\ref{fig:subprojector} to extract the partial syndrome in a QEC cycle.

As a convenience choice, we can choose to group stabilizers pairwise as $\{S_{2i-1}, S_{2i}\}$, which allows us to define the two-element subprojectors $\bar{P}^{\textbf{2}}_{i}$
\begin{equation}
    \bar{P}^{\textbf{2}}_{i} = P_{2i-1} P_{2i}=\frac{1}{4}(I+S_{2i-1})(I+S_{2i}),
\end{equation}
and it's easy to see
\begin{equation}
    \bar{P } = \prod^{k/2}_i \bar{P}^{\textbf{2}}_{i}.
\end{equation}
In such a way, we can halve the amount of readouts while the implementation of the projector gates $e^{i\pi \bar{P}^{\textbf{2}}_i}$ is still not too complicated. For instance, the [[5, 1, 3]] code has four stabilizer generators
\begin{equation} 
\begin{split}
        &S_1 = ZXXZI \quad S_2 = IZXXZ \\ 
        &S_3 = ZIZXX \quad S_4 = XZIZX,       
\end{split}
\end{equation}
and we can just use a single readout to measure the projector
\begin{multline}
    \bar{P} = \frac{1}{16}(I+ZXXZI)(I+IZXXZ)\\(I+ZIZXX)(I+XZIZX).
\end{multline}
This would require 18 two-qubit gates (depth 12) and a five-qubit Toffoli gate in total to implement the projector gate. Rather if we measure the following two subprojectors
 \begin{equation}
 \begin{split}
        &\bar{P}_1 = \frac{1}{4} (I+ZXXZI)(I+IZXXZ) \qquad \\ &\bar{P}_2 = \frac{1}{4} (I+ZIZXX)(I+XZIZX), 
 \end{split}
 \end{equation}
, we only need 12 two-qubit gates (depth 6) and a Toffoli gate for each subprojector gate and the overall gate overhead is 24 two-qubit gates (depth 12) and two Toffoli gates.

\begin{figure*}[t]
    \centering
    \includegraphics[width=.8\linewidth]{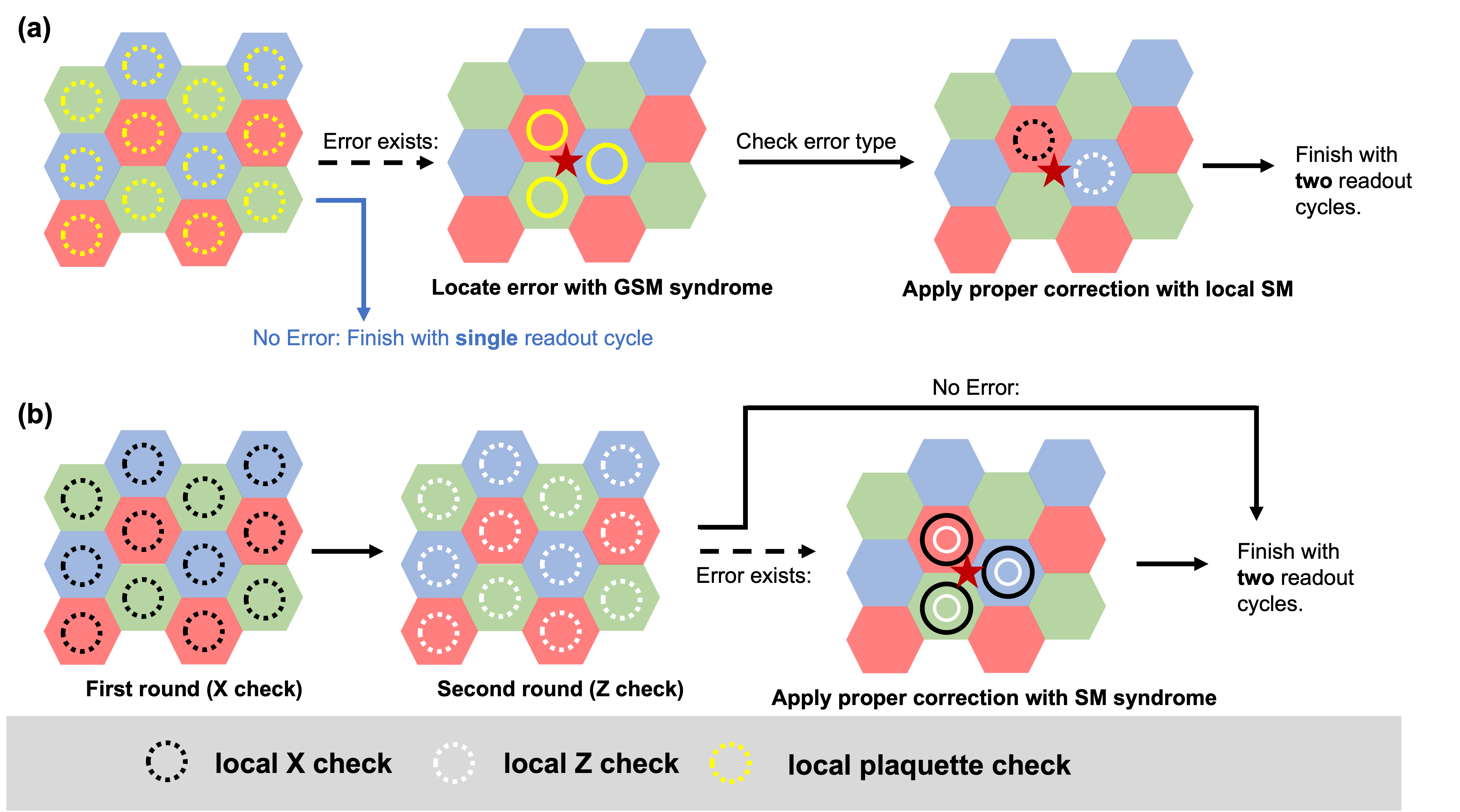}
    \caption{Adoption of the GSM method on scalable two-dimensional color codes (a) a QEC cycle with the GSM method. Yellow dashed circles represent for the local GSM routine on each plaquette. When error rate is low, we will mostly end with just a single-shot readout time. We follow with local stabilizer check when the location of error(s) is detected to correct the error(s), and the time consumed is two-shot readout time. (b) a canonical QEC cycle without the GSM method. Whether an error is detected or not, the time required is always two-shot readout time. }
    \label{fig:color_code}
\end{figure*}

Importantly, the partial syndrome information extracted with the GSM method can be further used in the correction stage in the QEC cycle. We only need to perform canonical syndrome measurement for stabilizers associated with triggered subprojectors, and stabilizers associated with untriggered subprojectors doesn't need to be measured again. Taking the two-shot GSM method on the $[[5, 1, 3]]$ code as an example: If only the second subprojector $\bar{P}_2$ is triggered, we only need to perform canonical syndrome measurement for stabilizer $ZIZXX$ and $XZIZX$, and there is no need to measure $ZXXZI$ and $IZXXZ$ again as we already assured that the logical state is in the space of subprojector $\bar{P}_1$. 

For another example, the two-dimensional color code \cite{fowler_two-dimensional_2011} has stabilizer $\{X^{\otimes n}, Z^{\otimes n} \}$ supported on each $n$-edge plaquette. We can use a subprojector associated with each plaquette 
\begin{equation}
    \bar{P}_{plaq} = \frac{1}{4}(I+X^{\otimes n})(I+Z^{\otimes n})
\end{equation}
to extract partial syndrome information (Fig.~ \ref{fig:color_code}).  For this kind of code, a QEC cycle with our GSM method only takes a single readout cycle when no error occurs, and at most two readout cycles in total when at least an error occurs. To be compared with, a canonical QEC cycle without the GSM method always takes two readout cycles. Therefore, our GSM method can always reduce the time of readout for two-dimensional color code, providing a scalable option for fault-tolerant quantum computing.

In general, there is no restriction on how to group the stabilizer generators into subprojectors, and the subprojectors can be chosen to mostly suit the connectivity and practical limitation of quantum hardware. The number of readouts can be chosen smoothly from 1, corresponding to the one-shot GSM method proposed in the main text with the largest gate overhead (minimum readout overhead), to $k$, corresponding to the canonical syndrome measurement with the minimum gate overhead (largest readout overhead).

\section{Steane code simulation with state-of-the-art noise parameter}
\label{app-stateofart}
\begin{figure*}[ht]
    \centering
    \includegraphics[width=.6\linewidth]{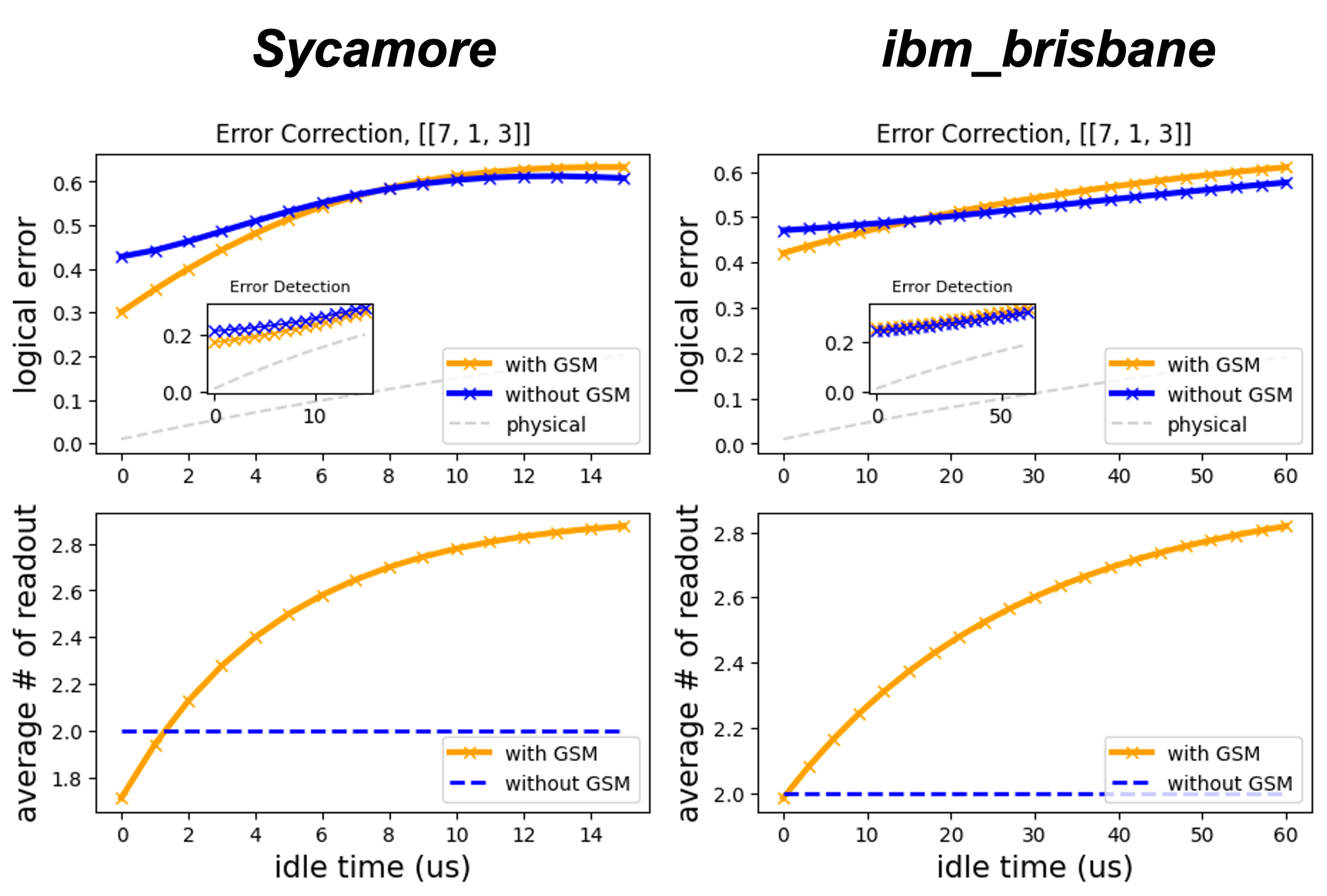}
    \caption{Steane code simulation with state-of-the-art noise parameter. }
    \label{fig:7_1_3}
\end{figure*}

In the main text, we present numerical results for the GSM method on the Steane code with projective noise parameters. The main reason why we don't stick with state-of-the-art noise parameter is the current superconducting quantum hardware is still too noisy and not in the regime where logical qubit would outperform physical qubit, as shown in Fig.~\ref{fig:7_1_3}. The logical error rate after the encoding and decoding is much larger than the physical error rate, which makes QEC not really meaningful in such a regime. However, we still see some advantage of the GSM method over the canonical SM method when the input logical state doesn't suffer too long idling time. As longer idling noise is applied to the logical state, the average number of readout of a QEC cycle with the GSM method increases over two (canonical SM method) and the GSM method start to lose its advantage. As raw physical qubits dominates over the logical qubits, this regime doesn't post a critique to our proposed GSM method.

\section{benchmark the advantageous regime of the GSM method}
\label{app-parachoice}
\begin{figure*}[]
    \centering
    \includegraphics[width=.9\linewidth]{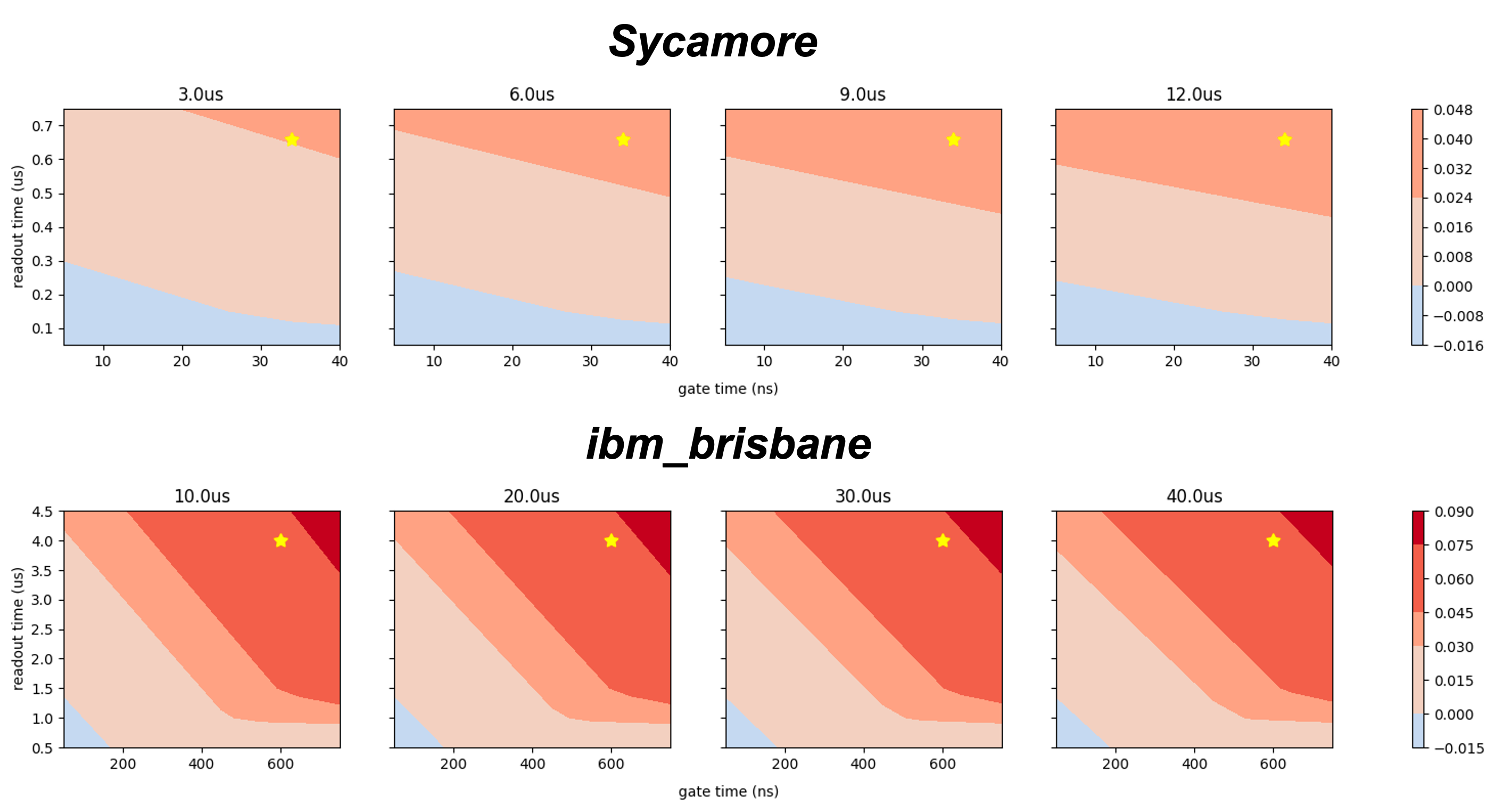}
    \caption{Fidelity difference $F_{gsm}-F_{sm}$ with $[[4, 2, 2]]$ code depending on various readout time and gate time when other noise parameters are fixed. The yellow stars denote where the state-of-the-art operation time is. }
    \label{fig:4_2_2}
\end{figure*}

\begin{figure*}[]
    \centering
    \includegraphics[width=.9\linewidth]{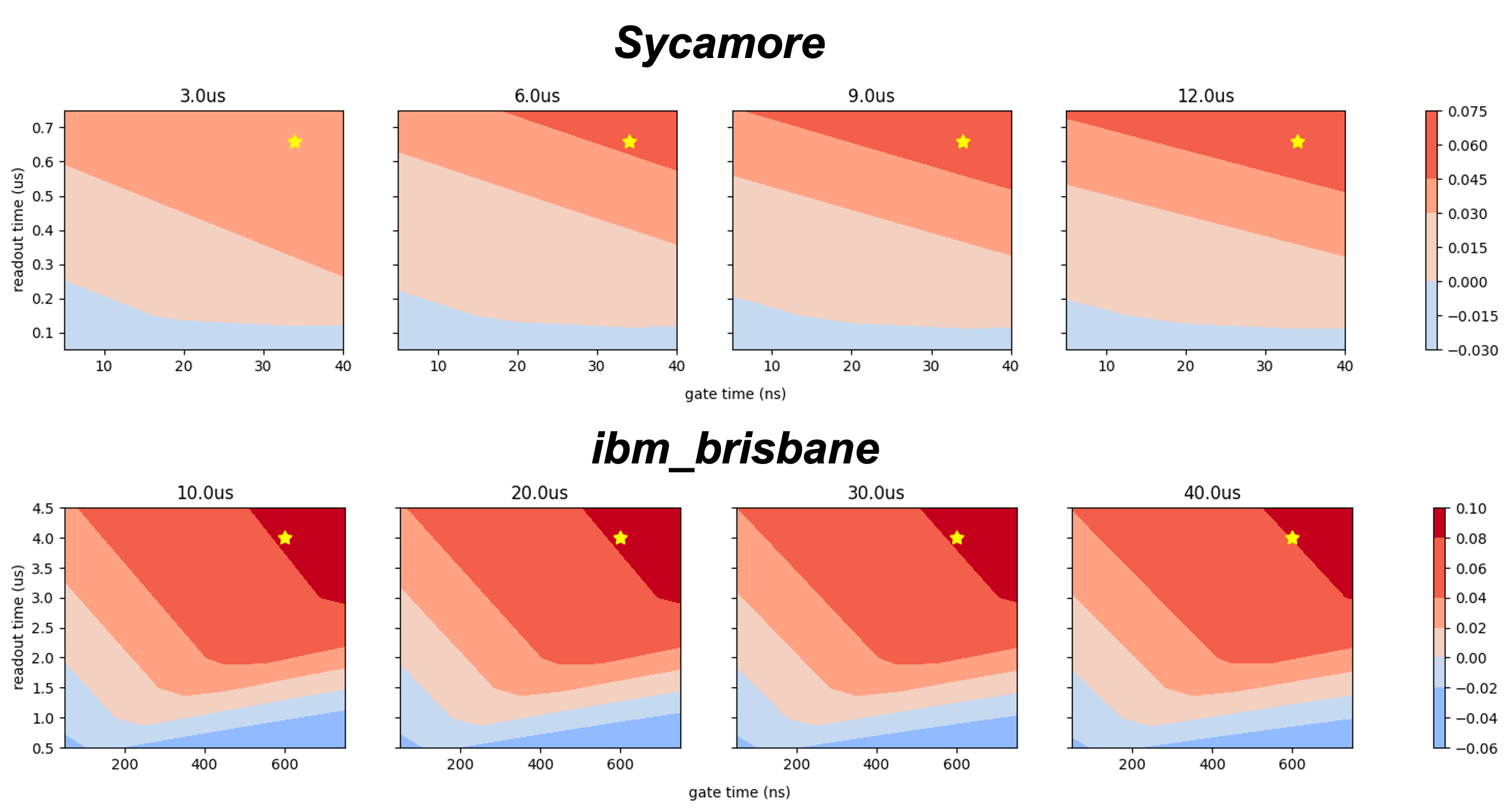}
    \caption{Fidelity difference $F_{gsm}-F_{sm}$ with $[[6, 4, 2]]$ code depending on various readout time and gate time when other noise parameters are fixed. The yellow stars denote where the state-of-the-art operation time is.}
    \label{fig:6_4_2}
\end{figure*}

To understand the advantageous regime of the GSM method, we ramp both the readout time and the gate time for both the \textit{Sycamore} and the \textit{ibm\_brisbane} setting within a practical range that might be reached in the future, while keep other noise parameters fixed. We compare the performance for both the GSM and the canonical SM method and plotted the fidelity difference $F_{gsm}-F_{sm}$ regarding the gate time and the readout time. We simulated both the $[[4, 2, 2]]$ code in Fig.~\ref{fig:4_2_2} and the $[[6, 4, 2]]$ code in Fig.~\ref{fig:6_4_2} to explore the condition when the GSM method is advantageous. The advantageous regime for the GSM(SM) method is plotted with red(blue) area. From the simulated results, we see the state-of-the-art parameter is in a range where the GSM method will beat the canonical SM method. Unless the readout duration is significantly reduced in the scalable quantum hardware, we will still stay in the advantageous regime for the GSM method. 

\section{Projector gate implementation for Iceberg code}
\label{appendix-gate}
\begin{figure}[h]
    \centering
    \includegraphics[width=\linewidth]{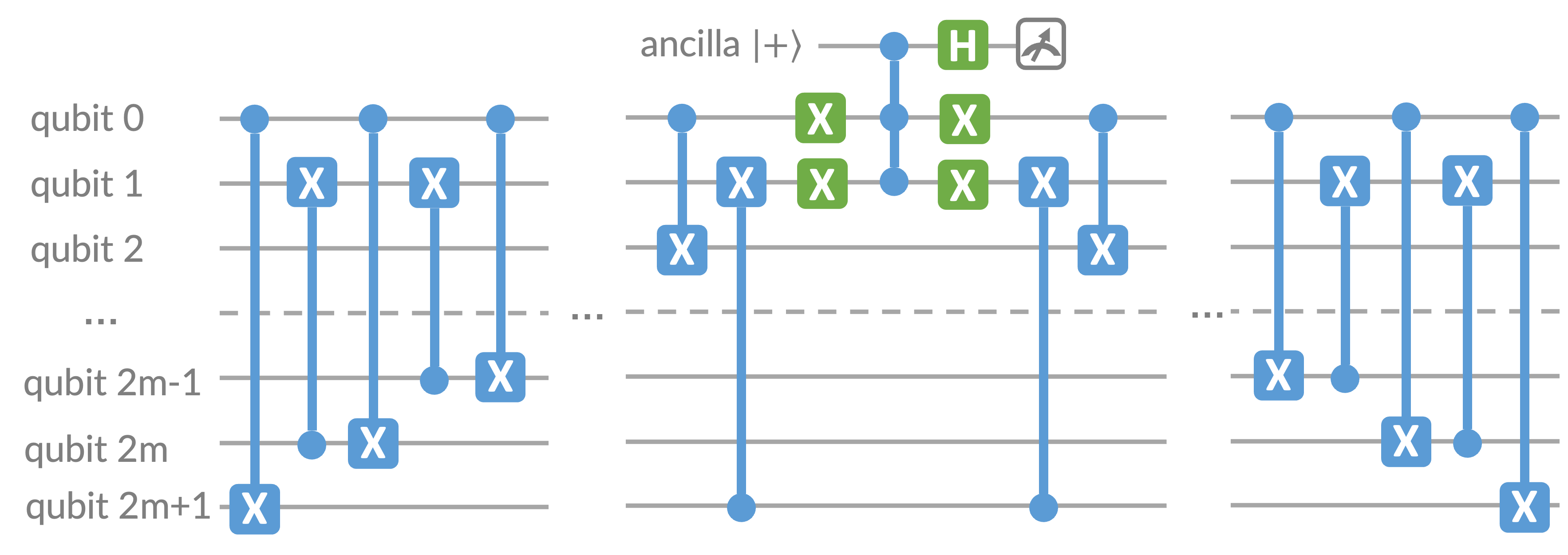}
    \caption{the projector gate implementation of for $[[2m+2, 2m, 2]]$ Iceberg code.}
    \label{fig:iceberg_gate}
\end{figure}

Although the implementation of the projector gate is not unique, here we present an implementation for general Iceberg code in Fig.~\ref{fig:iceberg_gate}. In such an implementation, the 2Q gate depth is only $4m$. Comparatively, the overall 2Q depth of the canonical syndrome measurement is $4m+4$.

\section{Fault-tolerance in analog with Shor-style syndrome measurement}
\label{app-faulttolerance}

In the long term, we have to consider the gate-level fault tolerance of our method to avoid the impact of correlated errors. To achieve this in our method, we require the quantum hardware to be able to implement native multi-qubit (control) Pauli phase gates and does not rely on decomposition into elementary gates. It's reasonable to make such an assumption as various quantum hardware has demonstrated potential for native multi-qubit gates \cite{levine_parallel_2019, kim_high-fidelity_2022}. We also demand that the native global gates don't suffer from inner correlated errors. As such, we only need to consider single Pauli errors induced by the gate operations up to the main order. 

\begin{figure}[h]
    \centering
    \includegraphics[width= \linewidth]{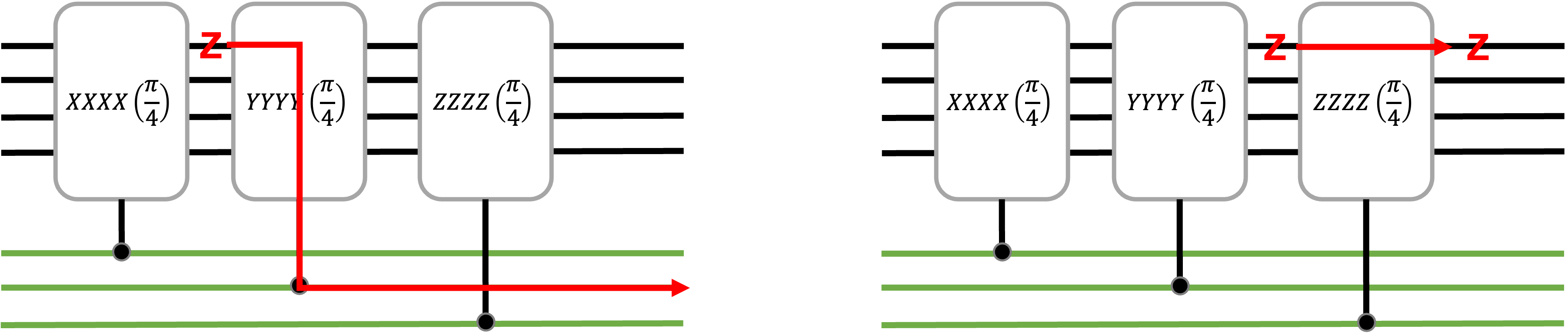}
    \caption{Generalized Syndrome Measurement in analog with Shor-style syndrome measurement for [[4, 2, 2]] code. Ancilla qubits (green) are prepared in the GHZ state $(\ket{000}+\ket{111})/\sqrt{2}$ and each ancilla qubit controls a multi-qubit Pauli gate. An error occurred after the gates will either be detected (left) or commute with the following gates and will not propagate (right).}
    \label{fig:FT}
\end{figure}

By using the idea of Shor-style syndrome measurement, i.e. preparing entangled ancilla states, we can achieve a fault-tolerant circuit for our method. We plotted the GSM circuit for [4, 2, 2] code as an example in Fig.~\ref{fig:FT}. The entangled ancilla states can prevent errors occurred in any single ancilla qubit. For errors occurred after the multi-qubit Pauli gates, they will either be detected by the following Pauli gates or commute with the following Pauli gates and will not be propagated to other qubits (Fig.~\ref{fig:FT}). 

Importantly, we note this fault tolerance cannot be achieved if we decompose the gate into elementary gates (two-qubit gates). That's because each Pauli phase gate is mediating global interaction and an error occurred after a decomposed gate could propagate to every qubit at the end of the Pauli phase gate. In early fault-tolerant setting, this may not be a severe problem if the gain from the reduced readout-induced noise is good enough, as shown in our simulation results.

\section{Quantum Error Correction with logical erasure conversion}
\label{app-logicalerasure}

\begin{figure}[h]
    \centering
    \includegraphics[width=\linewidth]{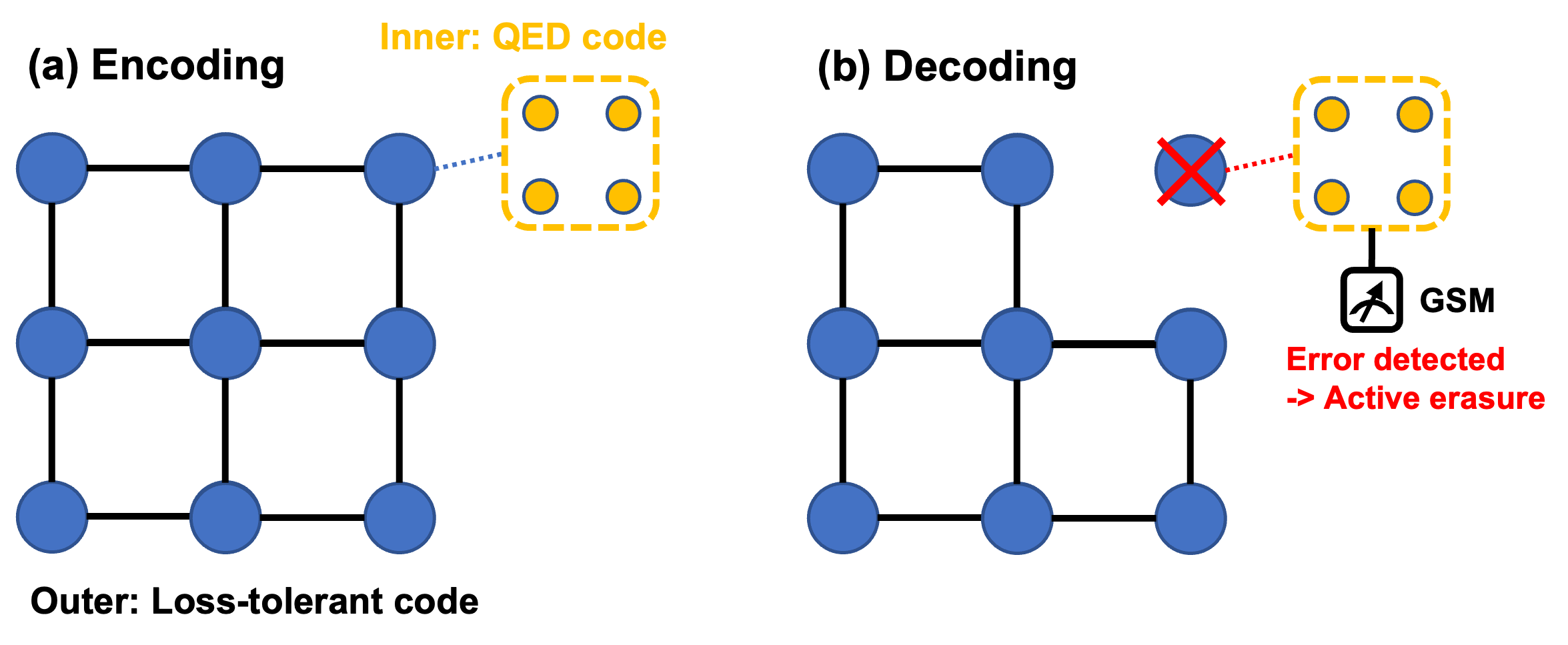}
    \caption{Scheme for concatenated codes with logical erasure conversion. In such a scenario, the ability to detect the errors is sufficient to correct physical errors. }
    \label{fig:concatenation}
\end{figure}

We can exploit the benefit of concatenated quantum codes to correct the errors indirectly (Fig.~\ref{fig:concatenation}). The idea is initially proposed in \cite{li_concatenation_2023} for cluster state generations. The inner codes are set to be small-size quantum code used to detect the physical errors, and the outer codes are used to correct the loss (erasure) errors. Whenever we detect an error in the inner level, we actively treat it as a logical erasure error in the outer level. As outer codes are loss-tolerant, we can restore the logical information in the presence of logical erasures and thus correct the physical errors indirectly. We do not need the syndrome information and correct the Pauli errors directly, and the GSM method can help reduce the readout-induced infidelity in such a scenario.

\end{document}